# Acceptable solutions of the Schrödinger radial equation for a particle in a two-dimensional central potential

J. Etxebarria

*Department of Physics, University of the Basque Country, UPV/EHU, Bilbao, Spain*

Email address:j.etxeba@ehu.es

**Abstract**

The stationary states of a particle in a central potential are usually taken as a product of an angular part $\Phi$ and a radial part $R$. The function $R$ satisfies the so-called radial equation and is usually solved by demanding $R$ to be finite at the origin. In this work we examine the reason for this requirement in the case of a two-dimensional (2D) central force problem. In contrast to some claims commonly accepted, the reason for discarding solutions with divergent $R(0)$ is not the need to have a normalizable wave function. In fact some wave functions can be normalized even if $R$ is singular at the origin. Instead, here we show that if $R$ is singular, the complete wave function $\psi=\Phi R$ fails to satisfy the full Schrödinger equation, but follows a equation similar to Schrödinger's but with an additional term containing the 2D Dirac delta function or its derivatives. Thus, $\psi$ is not a true eigenfunction of the Hamiltonian. In contrast, there are no additional terms in the equation for wave functions $\psi$ built from solutions $R$ that remain finite at the origin. A similar situation also occurs for 3D central potentials as has been shown recently. A comparison between the 2D and 3D cases is carried out.

# 1. Introduction

The central force problem in three dimensions (3D) is one of the most frequently treated examples in introductory quantum mechanics courses, given its enormous importance, especially in the field of atomic physics. In comparison, the case of central potentials in 2D is much less studied, although this problem also has its own interest in examples of modern science such as the electronic stationary waves in quantum corrals [1,2], the study of quantum chaos in the stadium billiards problem [3,4], or the analysis of 2D atomic models [5-7].

The stationary states $\psi$ in the 2D central force problem can be calculated without difficulty following the method of separation of variables. A straightforward procedure easily leads us to write $\psi$ as a product of an angular part $e^{im\varphi}$ depending on the azimuthal angle $\varphi$, with $m$ integer, and a radial part $R(r)$ that is obtained as a solution of the so-called radial equation [7,8],

$$-\frac{\hbar^2}{2\mu}\left[\frac{d^2 R(r)}{dr^2} + \frac{1}{r}\frac{dR(r)}{dr}\right] + \left[V(r) + \frac{\hbar^2 m^2}{2\mu r^2}\right]R(r) = ER(r) , \qquad (1)$$

where $r$ is the radial coordinate, $E$ the energy, $\mu$ the mass of the particle, and $m$ the azimuthal quantum number, $m=0, \pm 1, \pm 2, \ldots$

Equation (1) is usually solved with the additional condition that $R$ must be finite at the origin,

$R(0)<\infty$ \hfill (2)

In the literature [2-6,8-10], this requirement is often taken for granted without any explanation, or sometimes justified in terms of the wave function normalization condition. However, the wave function can be normalized even if $R$ diverges at the origin. As a famous example of a wave function infinite at the origin but physically acceptable we have the ground state of the hydrogen atom when it is studied by the Dirac equation [11]. On the other hand, there are non-normalizable wave functions that are perfectly acceptable if the stationary states to be calculated are scattering states.

In this paper we will see that expression (2) is indeed correct, but the reasons for its justification are different from those given in the literature. We will show that if $R(r)$ satisfies (1) and in addition $R(0)= \infty$, then the wave function $\psi = e^{im\varphi} R(r)$ does not satisfy the complete Schrödinger equation. A situation similar to this also occurs in 3D central force problems, as has been recently analyzed [12,13]. The mathematical reason for this behavior is that the usual expression for the Laplacian operator in polar coordinates

$$\nabla^2 = \frac{\partial^2}{\partial r^2} + \frac{1}{r}\frac{\partial}{\partial r} + \frac{1}{r^2}\frac{\partial^2}{\partial \varphi^2} \qquad (3)$$

is not valid for $r=0$. An analogous issue takes place with the corresponding 3D Laplacian expressed in spherical coordinates [14,15].

More explicitly, we will see that if $R$ satisfies (1) and diverges at the origin, then $\psi$ satisfies an equation similar to Schrödinger's but with an additional term, which contains the Dirac delta or its derivatives. In other words, this kind of $\psi$ is not a true solution of the Schrödinger equation.

The organization of this paper is as follows. In Section 2 we will study the asymptotic behavior of the solutions of (1) near $r=0$ as a previous step for the analysis of the terms that involve the Dirac delta. For states with zero angular momentum ($m=0$) the argument can be easily understood by undergraduate students of physics. This is the most important case and will be presented in Section 3. In Section 4 we will deal with states with non-zero angular momentum ($m \neq 0$) that are not normalizable. The analysis of this situation is more complicated and requires the use of distribution theory [14,16]. The calculations of this part are given in the Supplementary Material and can be omitted on a first reading. Finally in Section 5, we will made a comparison between the conditions that functions $R(r)$ must verify at the origin to be acceptable in 2D and 3D central force problems.

**2. Asymptotic behavior of the solutions of the radial equation at the origin**

Our study will be restricted to the so-called regular potentials for which $\lim r^2 V(r) \to 0$ when $r \to 0$. In these conditions, the behavior near the origin of the solutions of (1) is given by

$$R(r) \propto r^s, \text{ with } s = \pm |m|. \tag{4}$$

This asymptotic behavior is easily obtained by substituting the dependence $r^s$ in (1) and equating to zero the coefficient of the dominant term. As can be seen, except if $m=0$, there are 2 different asymptotic behaviors that correspond to the 2 linearly independent solutions of (1). The $m=0$ case is special, because Eq. (4) only gives the behavior of one of the two solutions. However, it can be checked by explicit substitution that the second solution has a logarithmic asymptotic behavior, i.e.,

$$R(r) \propto \ln kr \tag{5}$$

where $k > 0$ is a constant with dimensions of reciprocal length.

The solutions of (1) that behave near $r=0$ like $R(r) \propto r^{|m|}$ automatically satisfy condition (2), whereas the solutions with a behavior $R(r) \propto r^{-|m|}$ ($m \neq 0$), and solution (5) ($m=0$) do not satisfy (2). We will show below that the second set of solutions is unacceptable. We will treat the cases $m=0$ and $m \neq 0$, separately.

**3. Case $m=0$**

It is interesting to note that if $m=0$, even though $R(0)$ diverges in one of the 2 solutions of (1) [the one that behaves according to (5)], this does not imply necessarily that such a solution is non-normalizable, and therefore cannot be discarded on this reason.

An example will clarify this point. We will take as a potential $V(r)$ the case of the infinite circular well of radius $a$. For $r<a$ we have $V(r)=0$, and Eq. (1) becomes the Bessel equation of order 0, i.e.,

$$x^2 R''(x) + x R'(x) + x^2 R(x) = 0 , \qquad (6)$$

where the primes denote derivatives with respect to the dimensionless variable $x = r\sqrt{2\mu E}/\hbar$ (the allowed energies $E$ are positive). The solution of (6) is a linear combination of the Bessel and Neumann functions of order 0, $J_0(x)$ and $Y_0(x)$, which present at $x=0$ an asymptotic behavior like $x^0$ and $\ln x$ (except for a constant) respectively, as expected. We will analyze the second solution in more detail.

Figure 1 shows the function $xY_0^2$ in the interval $0< r<a$, which is proportional to the density of probability of presence in that region. As can be seen, there are no problems at the origin, and any integral of the type $\int_0^b xY_0^2(x)\,dx$, with $b$ finite, does not diverge.

For *r>a*, *V(r)*=∞, and the wave function must vanish. Then, the *x* value corresponding to *r=a* must coincide with $x_n$, one of the zeros of $Y_0$, which is finite (the example of the figure corresponds to the second zero $x_2$). The squared norm of the wave function is an integral of the above type with $b = x_n$, which is also finite. (In its turn, the requirement of continuity for the wave function at *r=a* allows us to deduce quantized values for the energy $E_n$ through the expression $x_n = a\sqrt{2\mu E_n}/\hbar$, but here this is secondary).

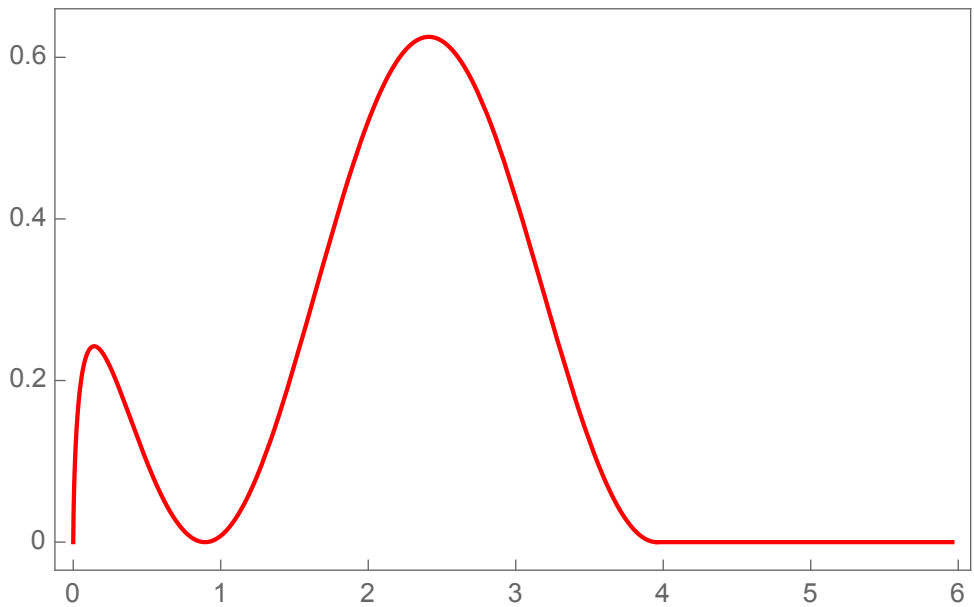

**Figure 1**. Density of probability of presence for a particle in an infinite circular well of radius *a* with wave function $Y_0(x)$ in the interval $0 < x < x_2$ ($x_2$ =3.95768 is the second zero of $Y_0$), corresponding to an energy $E_2 = \hbar^2 x_2^2 /(2\mu a^2)$. The probability of presence is null for $x > x_2$.

However, this solution is not acceptable because, although it satisfies (6), surprisingly, does not satisfy the Schrödinger equation! (At this point care should be taken not to identify (6) with the Schrödinger equation. Eq. (6) is nothing but the radial part of the Schrödinger equation, but it has a problem at the origin because it was derived from the full Schrodinger equation by assuming expression (3) for the Laplacian, *which is not valid* at *r*=0). In general, we will show below that for all the solutions of the radial equation (1) with $m=0$ and asymptotic behavior (5), an alternative expression to Schrödinger's equation results, of the type

$$\left[-\frac{\hbar^2}{2\mu}\nabla^2 + V(r)\right]R(r) = ER(r) + Q \quad , \tag{7}$$

where $Q$ is a term proportional to the 2D Dirac delta at the origin $\delta^{(2)}(\mathbf{r})$.

In order to verify (7) we will calculate the Laplacian of a function $F(r) = \ln kr$. To avoid singularities we will use a regularization procedure [17], taking $F(r)$ as the limit of $F_\alpha(r) = \ln[k(r+\alpha)]$ when $\alpha \to 0$. Since there is no angular dependence, we easily arrive at $\nabla^2 F_\alpha(r) = \left(\frac{\partial^2}{\partial r^2} + \frac{1}{r}\frac{\partial}{\partial r}\right)F_\alpha(r) = \frac{\alpha}{r(r+\alpha)^2}$, so $\nabla^2 F = 0$ if $r \neq 0$. For $r=0$, however, there is a singular behavior, and to find out its true value we will apply the Gauss theorem in 2D to the vector field $\nabla F_\alpha(r) = \hat{\mathbf{r}}/(r+\alpha)$, where $\hat{\mathbf{r}}$ is the unitary radial vector. According to the Gauss theorem, the divergence of a vector field integrated in a region $S_\varepsilon$ is related with the flux of that vector field through the boundary $C_\varepsilon$ of that region, i.e.,

$$\iint_{S_\varepsilon} \nabla \cdot (\nabla F_\alpha)\, dA = \int_{C_\varepsilon} \nabla F_\alpha \cdot \mathbf{n}\, ds \;,$$

where $dA$ is the infinitesimal element of area in $S_\varepsilon$, $ds$ the infinitesimal element of length along the curve $C_\varepsilon$, and $\hat{\mathbf{n}}$ is the unitary vector perpendicular to $ds$ directed outward. Taking as region $S_\varepsilon$ the circle centered at the origin with radius $\varepsilon$, and $C_\varepsilon$ the corresponding circumference (see Fig. 2), we have

$$\int_{C_\varepsilon} \nabla F_\alpha \cdot \mathbf{n}\, ds = \frac{2\pi\varepsilon}{\varepsilon + \alpha}$$

whose limit when $\alpha \to 0$ is $2\pi$. Therefore $\iint_{S_\varepsilon} \nabla \cdot (\nabla F) dA = \iint_{S_\varepsilon} \nabla^2 F dA = 2\pi$, and since

$\nabla^2 F = 0$ for $r \neq 0$, this implies that at the origin $\nabla^2 F$ must have the form of a Dirac delta, i.e.,

$$\nabla^2 F = 2\pi \delta^{(2)}(\mathbf{r}). \tag{8}$$

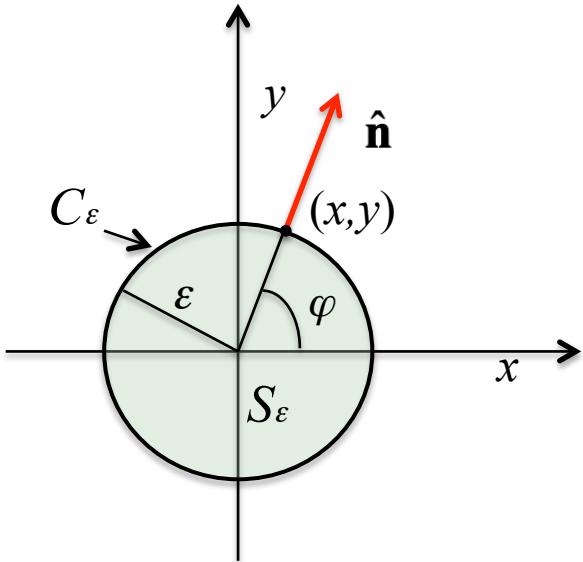

**Figure 2.** Circle $S_\varepsilon$ centered at the origin bounded by the circumference $C_\varepsilon$ of radius $\varepsilon$, and the normal unitary vector $\hat{\mathbf{n}}$ at an arbitrary point of $C_\varepsilon$ with coordinates $(x,y)$ or $(\varepsilon, \varphi)$.

Consequently, if in the case $m=0$ we take as a solution of (1) the function that approaches $R(r) = C \ln kr$ near $r=0$, with $C$ constant, there is a contribution $Q$ in (7) which is given by

$$Q = -\frac{\hbar^2 \pi}{\mu} C \delta^{(2)}(\mathbf{r}), \tag{9}$$

and such a function does not satisfy the Schrödinger equation.

This is the case of $Y_0(x)$ in the example of the circular well. Since we have

$$Y_0(x) \approx \frac{2}{\pi}\left(\gamma + \ln\frac{x}{2}\right)$$ near $x=0$ ($\gamma \approx 0.577$ is the so-called Euler constant), it is easy to find

$$-\frac{\hbar^2}{2\mu}\nabla^2 Y_0(r\sqrt{2\mu E}/\hbar) = E Y_0(r\sqrt{2\mu E}/\hbar) + Q, \tag{10}$$

where $Q = -\frac{2\hbar^2}{\mu}\delta^{(2)}(\mathbf{r})$.

It can be verified from Eq. (10) that the mean value of the Hamiltonian for a wave function proportional to $Y_0$ not only does not coincide with any of the eigenvalues of the Hamiltonian, but is divergent. Note also that the Hamiltonian is not Hermitian for the singular solution $Y_0$ as a consequence of satisfying (10) with $Q \neq 0$.

Evidently, none of these problems arise for the solution of (1) that behaves at the origin like $R(r) \propto r^0$ [$J_0(x)$ in the example of the circular well]. It is easy to see that in this case $Q=0$. This is the only valid solution for which, as can be seen, expression (2) is fulfilled.

## 4. Case $m \neq 0$

If $m \neq 0$, solutions of (1) that behave at the origin like $R(r) \propto r^{-|m|}$ are necessarily non-normalizable, since $\int_0^b r R(r)^2 dr$ diverges for any $b>0$ finite. This characteristic by itself already allows us to reject such solutions in bound state problems (although not necessarily in scattering problems). But we will also see that $\psi = e^{\pm i|m|\varphi} R(r)$ are not true eigenstates of the Hamiltonian $H$ either, but satisfy $H\psi = E\psi + Q$, with $Q \neq 0$.

The mathematics needed to understand this case is more advanced, requiring some knowledge of the theory of distributions [16]. The problems arise from the fact that the expression for the Laplacian $\Delta r^s = s^2 r^{s-2}$, that is always fulfilled for $r \neq 0$, is not valid at $r=0$ when $s$ is a negative even integer. In the latter case, there is an extra term involving derivatives of the Dirac delta [14]. The details can be seen in the Supplementary Material. In the following we will limit ourselves to give some results without demonstration.

Table 1 shows $Q$ for the first $m$ values in the case of functions $\psi = e^{\pm i|m|\varphi} R(r)$, with $R$ solution of (1) behaving like $R(r) \propto r^{-|m|}$ at $r=0$, so that we can write

$$\psi = e^{\pm i|m|\varphi} r^{-|m|} (a_0 + a_1 r + a_2 r^2 + ...), \qquad (11)$$

where $a_i$ are the coefficients of the Taylor series of $r^{|m|} R(r)$, and $a_0 \neq 0$.

| $m$ | $Q(-2\mu/\hbar^2)$ |
|---|---|
| $\pm 1$ | $2\pi a_0 (\partial_x \pm i\partial_y) \delta^{(2)}(\mathbf{r})$ |
| $\pm 2$ | $-\pi a_0 (\partial_{xx} - \partial_{yy} \pm 2i\partial_{xy}) \delta^{(2)}(\mathbf{r})$ |
| $\pm 3$ | $\pi (a_0/4)(-\partial_{xxx} - 2\partial_{yyx} \pm i\partial_{yyy} \pm 2i\partial_{xxy}) \delta^{(2)}(\mathbf{r})$ |

**Table 1** Extra contributions to the Schrödinger-type equation $H\psi = E\psi + Q$ satisfied by a wave function $\psi$ of the form (11)

As can be seen, all solutions of (1) that present a singularity of type $r^{-|m|}$ at the origin show an extra contribution $Q \neq 0$ and therefore functions of the form (11) must be ruled out as acceptable wave functions. Only the radial functions with asymptotic behavior $R(r) \propto r^{|m|}$ are acceptable. In all of them $R(0)=0$, thus satisfying expression (2).

**5. Comparison between the 2D and 3D cases. Concluding remarks**

As mentioned in the Introduction, a problem similar to the one discussed here also occurs when studying 3D central potentials. The two cases show similarities and differences.

In 3D the stationary states are also often taken as the product of an angular part (spherical harmonic) $Y_l^m$ (where $l=0,1,2,\ldots$, $m=-l,-l+1,\ldots 0,\ldots l-1,l$), and a radial part $R(r)$. The radial equation is different from (1), but there are also 2 linearly independent solutions, which behave near the origin like $r^l$ and $r^{-(l+1)}$ for the case of regular potentials. Similarly to what we have found in 2D, one of them gives rise to a physically acceptable state, while the other results in a wave function $\psi$ that does not satisfy the eigenvalue equation of the Hamiltonian $H\psi=E\psi$, but has an extra term $Q$ that involves the 3D Dirac delta at the origin $\delta^{(3)}(\mathbf{r})$ or its derivatives [12]. Table 2 summarizes schematically the situation.

**2D**

| $m$ | acceptable solution | unacceptable solution | $Q$ |
|---|---|---|---|
| 0 | $r^0$ | $\ln kr$ | $\delta^{(2)}(\mathbf{r})$ |
| ±1 | $r^1$ | $r^{-1}$ | $\partial_i \delta^{(2)}(\mathbf{r})$ |
| ±2 | $r^2$ | $r^{-2}$ | $\partial_{ij} \delta^{(2)}(\mathbf{r})$ |
| ±3 | $r^3$ | $r^{-3}$ | $\partial_{ijk} \delta^{(2)}(\mathbf{r})$ |

**3D**

| $l$ | $m$ | acceptable solution | unacceptable solution | $Q$ |
|---|---|---|---|---|
| 0 | 0 | $r^0$ | $r^{-1}$ | $\delta^{(3)}(\mathbf{r})$ |
| 1 | $0, \pm 1$ | $r^1$ | $r^{-2}$ | $\partial_i \delta^{(3)}(\mathbf{r})$ |
| 2 | $0, \pm 1, \pm 2$ | $r^2$ | $r^{-3}$ | $\partial_{ij} \delta^{(3)}(\mathbf{r})$ |
| 3 | $0, \pm 1, \pm 2, \pm 3$ | $r^3$ | $r^{-4}$ | $\partial_{ijk} \delta^{(3)}(\mathbf{r})$ |

**Table 2.** Behavior near the origin of the two independent solutions $R(r)$ of the radial equation in the 2D and 3D cases. The complete wave functions are $\psi = e^{im\varphi} R(r)$ and $\psi = Y_l^m R(r)$ respectively. For the acceptable solutions, the Schrödinger equation holds, whereas an anomalous term $Q$ appears in the equation for the unacceptable solutions.

As already mentioned, in the 2D case the acceptable solution is characterized by condition (2). Exactly the same condition must be required for the acceptable solutions in 3D, since the behavior of the non-acceptable solutions is also divergent at the origin ($R(r) \propto r^{-(l+1)}$) while the acceptable solutions ($R(r) \propto r^l$) have a finite value at $r=0$.

Both in 3D and 2D the radial equation can be written in an alternative form as a 1D Schrödinger equation in the variable $r$ for the function $u(r) = R(r)r$ (3D case) or $u(r) = R(r)\sqrt{r}$ (2D case). The corresponding Hamiltonian contains an effective potential $V_{eff}(r) = V(r) + l(l+1)\hbar^2/2\mu r^2$ (3D), or $V_{eff}(r) = V(r) + (m^2 - 1/4)\hbar^2/2\mu r^2$ (2D). In terms of function $u(r)$, condition (2) is equivalent in 3D to demand $u(0)=0$ (all acceptable solutions satisfy this equation and none of the unacceptable ones satisfy it). In fact, that expression is the most commonly used in the literature for 3D. However, in 2D there is no complete equivalence between (2) and $u(0)=0$. The exception is the solution with the logarithmic divergence for $m=0$, where $R(0)$ diverges but, however, $u(0)=0$.

In conclusion, in both 2D and 3D the radial equation must be solved with an additional condition at the origin to be physically acceptable. In both cases, failure to comply with this condition results in a wave function that does not satisfy the eigenvalue equation of the Hamiltonian, but has a singular contribution at $r=0$. The origin of the problem is that the usual mathematical expression for the Laplacian in polar or spherical coordinates is not valid at the origin. The requirement (2) allows the unambiguous differentiation of the acceptable solutions $R(r)$ both in 2D and 3D cases. The equality $u(0)=0$, which has the aspect of a boundary condition for the differential equation for $u(r)$, is equivalent to (2) in the 3D case but not so in 2D.

**Acknowledgements**

This work was supported by the Basque Country Government (Project No. IT1458-22)

# Supplementary Material

J. Etxebarria

*Department of Physics, University of the Basque Country, UPV/EHU, Bilbao, Spain*

**Calculations for singular solutions with *m*≠0**

If we omit the point *r*=0 it is immediate to show that the 2D Laplacian of $r^s$ is $\nabla^2 r^s = s^2 r^{s-2}$ (*r*≠0). However, distribution theory [1] shows that if we want to include the point *r*=0, the Laplacian has a different expression in the case where *s* is a negative even integer. More specifically, one has [1,2]

$$\nabla^2 r^s = s^2 r^{s-2} + \chi_p C_p \nabla^{2p} \delta, \text{ with } C_p = -\frac{4p\pi}{2^{2p-1}(p!)^2}, \ p = -\frac{s}{2}, \tag{1}$$

where $\nabla^{2p}\delta$ is the iterated Laplacian of the two-dimensional Dirac delta at the origin $\delta = \delta^{(2)}(\mathbf{r})$. The quantity $\chi_p$ is given by $\chi_p = 1$ if *p*=1,2,3,…, and $\chi_p = 0$ otherwise. It can be verified that the extra contribution occurs only for *s*=-2,-4,-6….

Expression (1) has important effects in the computation of $\nabla^2 \left( r^s e^{im\varphi} \right)$. Since $r^{|m|}e^{im\varphi}$ is a non-singular function (homogeneous polynomial of degree |*m*|), if we write $r^s e^{im\varphi}$ as a product $r^s e^{im\varphi} = r^{s-|m|} r^{|m|} e^{im\varphi}$, the usual relation of the Laplacian of a product of functions is verified [1-3]. Then we have

$$\nabla^2\left(r^s e^{im\varphi}\right) = \nabla^2\left(r^{s-|m|}\right)\left(r^{|m|}e^{im\varphi}\right) + \left(r^{s-|m|}\right)\nabla^2\left(r^{|m|}e^{im\varphi}\right) + 2\nabla\left(r^{s-|m|}\right)\cdot\nabla\left(r^{|m|}e^{im\varphi}\right) \ . \qquad (2)$$

On the other hand, we can easily show that $\nabla^2\left(r^{|m|}e^{im\varphi}\right) = 0$ and

$\nabla\left(r^{s-|m|}\right)\cdot\nabla\left(r^{|m|}e^{im\varphi}\right) = e^{im\varphi}|m|r^{|m|-2}\mathbf{r}\cdot\nabla\left(r^{s-|m|}\right)$, and thus Eq. (2) takes the form

$$\nabla^2\left(r^s e^{im\varphi}\right) = \nabla^2\left(r^{s-|m|}\right)\left(r^{|m|}e^{im\varphi}\right) + 2e^{im\varphi}|m|r^{|m|-2}\mathbf{r}\cdot\nabla\left(r^{s-|m|}\right) \ . \qquad (3)$$

The dot product of the last term also has singular contributions. A way to calculate them is from the expression

$$\nabla^2 r^{s-|m|+2} = r^{s-|m|}\nabla^2 r^2 + r^2\nabla^2 r^{s-|m|} + 2\nabla r^2\cdot\nabla r^{s-|m|} \ .$$

Using Eq. (3) and noting that $\nabla^2 r^2 = 4$ and $\nabla r^2 = 2\mathbf{r}$ , we obtain

$$\nabla^2\left(r^s e^{im\varphi}\right) = \left(1-\frac{|m|}{2}\right)r^{|m|}e^{im\varphi}\nabla^2 r^{s-|m|} + \frac{|m|}{2}r^{|m|-2}e^{im\varphi}\nabla^2 r^{s-|m|+2} - 2|m|e^{im\varphi}r^{s-2} \qquad (4)$$

Now, using Eq. (1) to calculate $\nabla^2 r^{s-|m|}$ and $\nabla^2 r^{s-|m|+2}$ we arrive at

$$\nabla^2\left(r^s e^{im\varphi}\right) = (s^2-m^2)e^{im\varphi}r^{s-2} + \left(1-\frac{|m|}{2}\right)r^{|m|}e^{im\varphi}\chi_p C_p \nabla^{2p}\delta + \frac{|m|}{2}r^{|m|-2}e^{im\varphi}\chi_{p-1}C_{p-1}\nabla^{2(p-1)}\delta \ ,$$

with $p = -\dfrac{s-|m|}{2}$ . \hfill (5)

A final simplification of (5) can be carried out using the relation $r^2\nabla^{2p}\delta = 4p^2\nabla^{2(p-1)}\delta$ ,

which is proved in the Supporting Material of ref. [4] (see Eq. (16) of that Supporting

Material). The final result is

$$\nabla^2 \left( r^s e^{im\varphi} \right) = (s^2 - m^2) e^{im\varphi} r^{s-2} + \left[ \left( 1 - \frac{|m|}{2} \right) \chi_p C_p + \frac{|m|}{8p^2} \chi_{p-1} C_{p-1} \right] e^{im\varphi} r^{|m|} \nabla^{2p} \delta ,$$

with $p = -\dfrac{s-|m|}{2}$, (6)

Eq. (6) will now be used to calculate the Laplacian of $\psi = e^{im\varphi} R(r)$, with $R(r)$ behaving like $1/r^{|m|}$ near the origin ($m = \pm 1, \pm 2, \pm 3, ...$). This behavior implies that the wave function can be expanded in series as

$$\psi = \frac{e^{im\varphi}}{r^{|m|}} \sum_{k=0}^{\infty} a_k r^k ,$$ (7)

where $a_k$ are appropriate coefficients.

We will consider the cases $|m|= 1, 2$ and $3$ as an illustration. A direct application of (6) gives the following results:

- For $|m|=1$ the only anomalous contribution in the Laplacian of the different terms of sum (7) comes from $e^{\pm i\varphi}/r$, which gives $\nabla^2 \left( e^{\pm i\varphi} a_0 / r \right) = -\pi e^{\pm i\varphi} a_0 r \nabla^2 \delta$. This result can be written in terms of derivatives of $\delta$. Using the properties $x \dfrac{\partial^n \delta}{\partial x^n} = -n \dfrac{\partial^{n-1} \delta}{\partial x^{n-1}}$ and

$x \dfrac{\partial^n \delta}{\partial y^n} = 0$ ($n$ positive integer), we have $e^{\pm i\varphi} r \nabla^2 \delta = (x \pm iy) \left( \dfrac{\partial^2 \delta}{\partial x^2} + \dfrac{\partial^2 \delta}{\partial y^2} \right) = -2 \left( \partial_x \pm i \partial_y \right) \delta$,

and then

$$\nabla^2 \left( e^{\pm i\varphi} a_0 / r \right) = 2\pi a_0 \left( \partial_x \pm i \partial_y \right) \delta$$ (8)

- For $|m|=2$ there is also a unique non-zero anomalous term in the Laplacian, which comes from $e^{\pm i 2\varphi}/r^2$. We have $\nabla^2 \left( e^{\pm i 2\varphi} a_0 / r^2 \right) = -\dfrac{\pi}{8} a_0 e^{\pm i 2\varphi} r^2 \nabla^4 \delta$, which can be transformed into an expression involving the second derivatives of the Dirac delta. After some algebra we obtain

$$\nabla^2 \left( e^{\pm i2\varphi} a_0 / r^2 \right) = -\pi a_0 \left( \partial_{xx} - \partial_{yy} \pm 2i\partial_{xy} \right) \delta . \tag{9}$$

- For |m|=3 there is also a single term with anomalous contribution, $e^{\pm i3\varphi} / r^3$. An analogous (but longer) procedure results in $\nabla^2 \left( e^{\pm i3\varphi} a_0 / r^3 \right) = -\dfrac{\pi}{192} a_0 e^{\pm i3\varphi} r^3 \nabla^6 \delta$, which leads finally to

$$\nabla^2 \left( e^{\pm i3\varphi} a_0 / r^3 \right) = -\dfrac{\pi}{2} a_0 \left( \partial_{yyx} \mp i\partial_{xxy} \right) \delta - \dfrac{\pi}{4} a_0 \left( \partial_{xxx} \mp i\partial_{yyy} \right) \delta . \tag{10}$$

As can be seen, all these extra contributions can be written in terms of the successive partial derivatives of δ.